\documentclass[aps,prd,floats,cite,preprint,tightenlines,nofootinbib]{revtex4}

\usepackage{graphicx}

\newcommand{\beq}{\begin{eqnarray}}% can be used as {equation} or {eqnarray}
\newcommand{\eeq}{\end{eqnarray}}

\begin{document}
 
\preprint{ \hbox{hep-ph/0305302} }
\vspace*{3cm}
\title{A Simple Model of two Little Higgses}
\author{Witold Skiba\footnote{email address:  {\tt witold.skiba@yale.edu}} 
             and John Terning\footnote{email address:  {\tt terning@lanl.gov}} }
\affiliation{\small $^*$Department of Physics, Yale University, New Haven, CT 06520\\
$^\dagger$Theory Division T-8, Los Alamos National Laboratory, Los Alamos, NM  87545 \vspace{2.5cm}}

%Create the title page

\begin{abstract}
We construct a little Higgs model using a simple global symmetry group $SU(9)$
spontaneously broken to $SU(8)$.  The electroweak interactions are extended 
to $SU(3)\times U(1)$ and embedded in $SU(9)$. At the electroweak scale,
our model is a two Higgs-doublet model. At the TeV scale, there are additional 
states, which are responsible for the cancellation of one loop quadratic divergences.
We compute the effects of heavy states on the precision electroweak observables
and find that the lower bounds on the masses of heavy gauge bosons and fermions are
between 1 and  2~TeV. 
 \end{abstract}
%%%%%%%%%%%%%%%%%%%%%%%%%%%%%%%%%%%%%%%%%%%%%%%%
%\tighten

\maketitle

\newpage
%%%%%%%%%%%%%%%%%%%%%%%%%%%%%%%%%%%%%%%%%%%%%%%%
%Main body of the paper

%%%%%%%%%%%%%%%%%%%%%%%%%%%%%%%%%%%%%%%%%%%%%%%%
\section{Introduction}
%%%%%%%%%%%%%%%%%%%%%%%%%%%%%%%%%%%%%%%%%%%%%%%%
There is little doubt that the Standard Model of electroweak interactions is an effective
theory valid below 1~TeV. New interactions will come into play around a TeV and resolve
the hierarchy problem, which would be present in the low-energy theory if one tried
to set the cutoff to a higher value. There are several scenarios for what the new physics
at the TeV scale might be: supersymmetry, extra dimensions, dynamical symmetry breaking.
Typically, weakly interacting theories at the TeV scale are in better agreement with the precision
measurements  of electroweak observables. 

Little Higgs theories~\cite{ACG2,twosite,ACKN,GW,SU6Sp6,KS,littlecustodial}
provide another weakly-coupled alternative for how the Standard Model could be embedded
in a theory valid beyond 1~TeV. The little Higgs models employ an extended set of global
and gauge symmetries in order to get rid of one-loop quadratic divergences. Without the
one-loop divergences a cutoff of about 10~TeV or so is natural for a Higgs theory.
While this may not seem like a great achievement on the road to a fundamental theory,
compared for example with supersymmetry, 10~TeV is a very special scale from a practical
viewpoint. The next generation of collider experiments will not be able to probe energies
beyond a few TeV. It would certainly be fascinating to uncover a theory valid all the way to
the Planck scale, but we will only have experimental data up to a few TeV. The energy scales
above 10~TeV may remain a subject of speculation for  a very long time. 

Fortunately, different classes of extensions of the Standard Model make different predictions
for  the TeV spectrum. New states must enter the theory no later than a TeV to avoid fine tuning. 
The cancellation of one-loop divergences in little Higgs theories
requires new, heavy states for every divergence that is numerically significant at one loop.
Thus, little Higgs theories predict a whole set of heavy states: vector bosons, fermions,
and scalars to cancel the diagrams involving Standard Model gauge bosons, top quark,
and the Higgs; respectively.  Given a little Higgs model, it is certainly possible to make
specific predictions since the theory is perturbative below the cutoff. Some more detailed studies have
been reported in Ref.~\cite{ACGW,CES,littlepheno}. 

Little Higgs theories draw on the old idea that the Higgs could be a pseudo-Goldstone
boson~\cite{composite1,composite2}. The important recent development is the understanding
of how to cleverly arrange the breaking of global symmetries. The breaking of symmetries
that protect the Higgs mass needs to be such that the Higgs quartic
coupling is of order one, the Higgs couples to fermions and gauge bosons,
and at the same time one avoids quadratic divergences~\cite{ACG2}. Symmetry breaking
by interactions with numerical coefficients of order one is arranged such that individual
symmetry-breaking terms do not introduce a Higgs mass. To get mass terms for the Higgs
one needs more than one insertion of symmetry breaking interactions, and 
such diagrams are not quadratically divergent at one loop.

In what follows we present a little Higgs theory based on a simple global symmetry group,
$SU(9)$ which is spontaneously broken to $SU(8)$.  The models in Refs.~\cite{ACKN,SU6Sp6}
also have simple global symmetries, but the breaking patterns are different.
Our model also has a simple\footnote{The definition of what constitutes a  simple group is different in little Higgs theories than it is in group theory textbooks.} electroweak gauge group -- $SU(3)\times U(1)$, analogous to Ref.~\cite{KS}. 
As in the model with an $SU(4)\times U(1)$  gauge group \cite{KS} there is no mixing between  
the light and heavy charged gauge bosons induced by the Higgs VEV in this theory. Our $SU(3)\times U(1)$ model has only one additional $U(1)$ compared to the Standard Model and thus only one $Z'$ gauge boson. The tree-level exchanges of the single $Z'$ result in corrections that are one half the size of those in the $SU(4)\times U(1)$ model \cite{KS,littleprecision2} and thus in comparison are relatively benign for
constraining the model by current data. 

In the next section we describe the model.  The theory is based on the interactions of the Goldstone bosons associated with $SU(9)\rightarrow SU(8)$ breaking.  We show how to embed 
$SU(3)\times U(1)$ in $SU(9)$. Our embedding produces two 
weak scalar doublets and two complex singlets. The model has no weak triplets.
We then show how to obtain an acceptable Higgs quartic coupling as well as Yukawa couplings.
Since the gauge symmetry is $SU(3)\times U(1)$ all left-handed fields need to come in triplets.
In Sec.~\ref{sec:pheno} we follow Refs.~\cite{littleprecision1,littleprecision2}
and compute the tree-level effects of the heavy states on the precision electroweak data.
The lower bound on the masses of the heavy states is around 2~TeV. 
%%%%%%%%%%%%%%%%%%%%%%%%%%%%%%%%%%%%%%%%%%%%%%%%
\section{The Model}
\label{sec:model}
%%%%%%%%%%%%%%%%%%%%%%%%%%%%%%%%%%%%%%%%%%%%%%%%
 Our model is based on the $SU(9)/SU(8)$ nonlinear sigma model, in which $SU(3)\times U(1)$
 gauge interactions are embedded. When $SU(9)$ global symmetry is broken to its $SU(8)$
 subgroup, the  $SU(3)\times U(1)$ group is broken to the electroweak $SU(2)_W\times U(1)_Y$
 group.
 
 The $SU(9)/SU(8)$ coset space is described by a field $\Sigma$, which transforms linearly
 $\Sigma \rightarrow L \Sigma$ under $L \in SU(9)$ transformations. $\Sigma$ can be
 expressed\footnote{The unusual normalization of $f$ is chosen to agree with that of the $SU(4)$ model  in  Refs. \cite{KS,littleprecision2} so as to ease the comparison of the two models.}  in terms of pion fields $\Sigma=\exp( i \hat{\pi}/\sqrt{2}f)\hat{v}$,
where 
 \begin{equation}
 \label{eq:pions}
   \hat{\pi}=\left( \begin{array}{cc|cc|cc}
              0 & 0 & 0 &  \frac{f_1}{f} h_1 & 0 & \frac{f_2}{f} h_1\\
              0 & 0 & 0 & \frac{f_1}{f} s_1 & 0 & \frac{f_2}{f} s_1 \\ \hline
              0 & 0 & 0 &\frac{f_1 f_2}{f^2} h_2 & 0 &\frac{f_2^2}{f^2} h_2 \\  
             \frac{f_1}{f} h_1^\dagger &\frac{f_1}{f} s_1^* &  h_2^\dagger & 
                - 2\frac{f_1 f_2}{f^2} s_2^R &  -\frac{f_2^1}{f^2} h_2^\dagger & 
                  \frac{f1^2-f_2^2}{f^2} s_2^R + i s_2^I  \\ \hline 
              0 & 0 & 0 & -\frac{f_2^1}{f^2} h_2 & 0 & -\frac{f_1 f_2}{f^2} h_2  \\  
               \frac{f_2}{f}  h_1^\dagger &  \frac{f_2}{f} s_1^* & \frac{f_2^2}{f^2}  h_2^\dagger & 
                  \frac{f1^2-f_2^2}{f^2} s_2^R - i s_2^I & 
                  -\frac{f_1 f_2}{f^2} h_2^\dagger & 2 \frac{f_1 f_2}{f^2} s_2^R  
            \end{array} \right)
    \; {\rm and} \; 
    \hat{v}= \left( \begin{array}{c} 0 \\ 0 \\ \hline  0 \\ f_1 \\ \hline  0 \\ f_2 \end{array} \right) .
\end{equation}
In the equation above, $s_i$ are electroweak singlet fields while $h_i$ are electroweak doublets.
For brevity, odd numbered rows and columns  are two dimensional, while even numbered ones
are one dimensional. Moreover, $f^2= f_1^2+ f_2^2$ and $s_2^R$, $s_2^I$ are the real and imaginary parts of singlet $s_2$, similarily $h_1=h_1^R+ i h_1^I$ etc.
 The particular choice of $\hat{v}$ is meaningful because the global $SU(9)$
symmetry is explicitly broken by gauge interactions, which are described below. Otherwise
we could rotate $\hat{v}$ to have a nonzero entry in only one of its components.

The pion matrix, $\hat{\pi}$, does not include all $17= 80-63$ fields parameterizing
the $SU(9)/SU(8)$ coset space. We chose to work in the unitary gauge, in
which 5 fields that become the longitudinal components of the heavy gauge bosons
are set to zero. The $SU(3)\times U(1)_X$ generators are chosen to be
\begin{equation}
\label{eq:gaugegenerators}
  T^a =\frac{1}{2} \left( \begin{array}{c|c|c}
          t^a & 0 & 0 \\ \hline 
          0 & t^a & 0 \\ \hline 
          0 & 0 & t^a \end{array} \right)  \; {\rm and} \;         
  X= \frac{1}{3}  \left( \begin{array}{c|c|c}
          1 & 0 & 0 \\ \hline 
          0 & 1 & 0 \\ \hline 
          0 & 0 & 1  \end{array} \right),
\end{equation}
where $t^a$ are the $SU(3)$ Gell-Mann matrices normalized such that ${\rm tr}(t^a t^b)=2 \delta^{ab}$. Each entry in Eq.~(\ref{eq:gaugegenerators}) denotes a 3 by 3 block. 
Therefore, under  the $SU(3)\times U(1)_X$ the sigma field decomposes as
$\Sigma \rightarrow {\bf 3}_{1/3} +  {\bf 3}_{1/3} +  {\bf  3}_{1/3}$.  This embedding of gauge
symmetry is reminiscent of the model in Ref.~\cite{Kosower}.
 The generators unbroken by $\hat{v}$ are $T^i$ for $i=1,2,3$ and $Y=\frac{-1}{\sqrt{3}} T^8-X$. Under the electroweak  group $SU(2)_W\times U(1)_Y$ the pion fields transform linearly, such that
$h_i$ transform as  ${\bf 2}_{-1/2}$ and  $s_i$ as  ${\bf 1}_{0}$. We use a nonstandard
hypercharge assignment for the Higgs doublets $h_i$, opposite to the usual one.
In our convention the doublets get VEVs in their upper components.

The transformation property of $\Sigma$,  $\Sigma \rightarrow L \Sigma$ , implies
that the covariant derivative of $\Sigma$ is
\begin{equation}
  D_\mu \Sigma = (\partial_\mu  + i g T^a A^a_\mu  + i g_X X X_\mu )\Sigma.
\end{equation}
In the equation above the $SU(3)$ gauge coupling is denoted as $g$ and it is equal,
at tree level,  to the $SU(2)_W$ coupling.
The normalization of fields in Eq.~(\ref{eq:pions}) is such that the kinetic
energy term is
\begin{equation}
\label{eq:kinetic}
  {\cal L}_{kin} = (D_\mu \Sigma)^\dagger D^\mu \Sigma.
 \end{equation}

Gauge interactions induce quadratically- and log-divergent terms as can be deduced
by computing the Coleman-Weinberg potential. The quadratically-divergent term
\begin{equation}
  g^2 \frac{\Lambda^2}{16 \pi^2} \Sigma^\dagger T^a T^a \Sigma
\end{equation}
is independent  of the pion fields since $T^a T^a$ is proportional to the identity. The same
is also true for the quadratically divergent contribution arising from the $U(1)_X$ interactions.
The log-divergent contribution
\begin{equation}
\label{eq:gaugelog}
  \frac{g^4}{16 \pi^2} \log(\frac{\Lambda^2}{g^2 f^2}) (\Sigma^\dagger T^a \Sigma)^2
\end{equation}
is small and generates masses for the pion fields of order $\frac{g^2}{4 \pi} f$,
which is of the order of the electroweak scale. Since the gauge interactions do not generate
any quadratically divergent terms these interactions are not sufficient for providing the 
Higgs quartic potential. 

\subsection{Quartic potential}
\label{sec:quartic}
%%%%%%%%%%%%%%%%%%%%%%%%%%%%%%%%%%%%%%%%%%%

The underlying technique of constructing little Higgs theories is to arrange interactions such 
that each individual interaction with an $\mathcal{O}(1)$ coefficient maintains enough global symmetry to ensure that all $SU(2)$  doublets are exact Goldstone bosons. Only several interactions acting together break enough symmetry and turn Higgs doublets into pseudo-Goldstone bosons. Interactions with small coefficients are numerically unimportant and can
have an arbitrary pattern of symmetry breaking.

In order to generate a quartic potential for the two Higgs doublets $h_i$ we explicitly
break the $SU(9)$ global symmetry. We add two terms
\begin{equation}
\label{eq:quartic}
  V_q= \kappa_1 (\Sigma^\dagger M_1 \Sigma)^2 +  
             \kappa_2 (\Sigma^\dagger M_2 \Sigma)^2,
\end{equation}
where
\begin{equation}
\label{eq:matricesM}
 M_1 = \left( \begin{array}{c|c|c}
          0 & 1 & 0 \\ \hline 
          1 & 0 & 0 \\ \hline 
          0 & 0 & 0  \end{array} \right)
   \; {\rm and} \;         
  M_2 = \left( \begin{array}{c|c|c}
          0 & 0 & 1 \\ \hline 
          0 & 0 & 0 \\ \hline 
          1 & 0 & 0  \end{array} \right).
\end{equation}
Each term separately preserves a different $SU(3)^3 $  subgroup of global $SU(9)$.
Neglecting gauge interactions, the $SU(3)^3$ symmetry is exact.
In both cases $SU(3)^3$ is broken to $SU(2)^3$ by the VEV $\hat{v}$. Such breaking
generates three sets of exact Goldstone bosons, which transform as doublets
under $SU(2)_W$. One linear combination of these three doublets is eaten when
the $SU(3)$ gauge group is broken to $SU(2)_W$. The remaining two  linear combinations
are two physical Higgs doublets, which stay exactly massless when one of the $\kappa_i$'s
is set to zero.

Note that this would not be the case for many other choices of symmetry
breaking spurions $M$. For example, spurion 
\begin{equation}
 \tilde{M}= \left( \begin{array}{c|c|c}
          1 & 0 & 0 \\ \hline 
          0 & 0 & 0 \\ \hline 
          0 & 0 & -1  \end{array} \right)
\end{equation}
also breaks $SU(9)$ to $SU(3)^3$. However, this particular $SU(3)^3$ subgroup is broken 
by $\hat{v}$ to $SU(3)\times SU(2)^2$ producing two massless doublets instead of three.

The two terms in Eq.~(\ref{eq:quartic}) acting together break $SU(9)$ to 
a single $SU(3)$. This preserved $SU(3)$ is the same as the gauged $SU(3)$ whose
generators are given in Eq.~(\ref{eq:gaugegenerators}). The breaking of $SU(3)$ to $SU(2)$ yields only one doublet, which is eaten.  Therefore,  the uneaten Higgs doublets become
pseudo-Goldstone bosons and can obtain non-derivative interactions.

Expanding $V_q$ to quadratic order in the singlet fields and to the quartic order in doublets
we obtain:
\begin{equation}
\label{eq:quarticcomp}
  V_q=2 \kappa_1 [ f_1 Im(s_1) - \frac{f_2}{\sqrt{2} f} Re(h_2^\dagger h_1)]^2 + 
            2 \kappa_2  [f_2 Im(s_1) + \frac{f_1}{\sqrt{2} f} Re(h_2^\dagger h_1) ]^2 + {\mathcal O}(s^3,h^5).
\end{equation}
Integrating out $Im(s_1)$ gives the quartic interaction
\begin{equation}
\label{eq:quarticfinal}
 V_{eff} =  \frac{ \kappa_1 \kappa_2 f^2}{\kappa_1 f_1^2+ \kappa_2 f_2^2}
    \left[ Re(h_2^\dagger h_1) \right]^2,
 \end{equation} 
which indeed vanishes when either $\kappa_1$ or $\kappa_2$ is zero.
  
The model as it is defined up to now needs to be amended, otherwise there is 
a potential problem. The omitted $ {\mathcal O}(s^3)$ terms in Eq.~(\ref{eq:quarticcomp})
include the term
\begin{equation}
\label{eq:tadpole}
  2 \sqrt{2}  (\kappa_2-\kappa_1) \frac{f_1 f_2}{f} Im(s_1)^2 s_2^I.
\end{equation} 
This terms generates a quadratically divergent tadpole for the field $s_2^I$, which
means that $s_2^I$ gets a VEV. The VEV for $s_2^I$ rotates between the nonzero
components of $\hat{v}$ in Eq.~(\ref{eq:pions}). If $\langle s_2^I \rangle$ is ${\mathcal O}(f)$ 
it will create a hierarchy between $f_1$ and $f_2$, that is either $f_1 \ll f_2$ or $f_1 \gg f_2$. 
We assumed so far in our analysis that $f_1$ and $f_2$ are of the same order of magnitude.

There are two simple ways of ensuring that $\langle s_2^I \rangle$ is small compared
to $f$. First, we can impose a discrete symmetry that interchanges the two terms in Eq.~(\ref{eq:quartic}). Then $\kappa_1=\kappa_2$ and the tadpole-generating term is absent.
Second, we can add  a potential which gives a mass to $s_2^I$. A linear term will force a VEV
for $s_2^I$, but a large enough mass term will prevent this VEV from being of order $f$.
A suitable potential is for example
\begin{eqnarray}
\label{eq:stabilize}
  V_s&=& \rho_1 (\Sigma^\dagger N_1 \Sigma)^2 +  
             \rho_2 (\Sigma^\dagger N_2 \Sigma)^2 \nonumber \\
           & =& 2 \sqrt{2}  s_2^I \frac{f_1 f_2}{f} (\rho_2 f_2^2- \rho_1f_1^2) + 
              (s_2^I)^2 (\rho_1 \frac{3 f_1^2 f_2^2 -f_2^4}{f^2} + 
                        \rho_2 \frac{3 f_1^2 f_2^2 -f_1^4}{f^2})- \nonumber \\
          & &  (\rho_2 f_2^2- \rho_1f_1^2) \left[ \frac{f_2^2}{f^2} h_1^\dagger h_1
                        + \frac{f_1^2-f_2^2}{f^2}  h_2^\dagger h_2 \right]
                       + \ldots,
\end{eqnarray}
where
\begin{equation}
 N_1 = \left( \begin{array}{c|c|c}
          1 & 0 & 0 \\ \hline 
          0 & 1 & 0 \\ \hline 
          0 & 0 & 0  \end{array} \right)
   \; {\rm and} \;         
  N_2 = \left( \begin{array}{c|c|c}
          0 & 0 & 0 \\ \hline 
          0 & 0 & 0 \\ \hline 
          0 & 0 & 1  \end{array} \right).
\end{equation}
Each term preserves an $SU(6)\times SU(3)$ subgroup of $SU(9)$, which is spontaneously broken to $SU(5)\times SU(2)$. Therefore, adding these two stabilizing terms does not introduce
a large Higgs mass.  It is clear by examining Eq.~(\ref{eq:stabilize}) that the mass terms for
the Higgs fields vanish when the tadpole term for $s_2^I$ vanishes. At the minimum of the
potential
there is no tadpole and a Higgs mass is not generated by the stabilizing potential.

\subsection{Yukawa couplings}
\label{sec:yukawa}
%%%%%%%%%%%%%%%%%%%%%%%%%%%%%%%%%%%%%%%%%%%%%%%%

Incorporating Yukawa couplings is  straightforward, and can be done in a manner outlined
in Ref.~\cite{KS}. Here we will show how to include the Yukawa couplings for the third family,
other families of quarks and leptons can  be included in the same way. Since the electroweak
gauge interactions are enlarged to $SU(3) \times U(1)$ all families of quarks and leptons
need to be extended to form representations of the larger electroweak group.
This is not necessary in most little Higgs models, in which the enlarged electroweak group contains an $SU(2) \times U(1)$ subgroup. In such cases, the light families can be coupled
directly to the Higgs doublet without regard to one-loop quadratic divergence, as this
divergence is numerically unimportant for the light fermions.

We add a pair of vector like fermions $\chi_L$ and $\hat{\chi}_R$ such that
$Q_L=(t_L,b_L,\hat{\chi}_L)^T$ transforms as an $SU(3)$ triplet. The $U(1)_X$ charges
of $Q_L$, $\hat{\chi}_R$,  $\hat{t}_R$, and $b_R$ are $-\frac{1}{3}$,  $-\frac{2}{3}$,  $-\frac{2}{3}$,  
and $\frac{1}{3}$, respectively.   
It is useful to introduce projection operators into $SU(3)$ subspaces of $\Sigma$:
\begin{equation}
  \Sigma_i = P_i \Sigma,
\end{equation}
where $P_1 = \left( \begin{array}{c|c|c} 1 & 0 & 0 \end{array} \right)$, 
$P_2 = \left( \begin{array}{c|c|c}  0 & 1 & 0 \end{array} \right)$, and
$P_3 = \left( \begin{array}{c|c|c}  0 & 0 & 1 \end{array} \right)$. Each entry in the
projection operators represents a three by three matrix. In terms of the 
three-vectors $\Sigma_i$, the Yukawa couplings  are generated by
\begin{equation}
\label{eq:yukawa}
 {\cal L}_{yuk}=\lambda_1\, \overline{Q}_L\, \Sigma_2 \, \hat{\chi}_R +
     \lambda_2 \, \overline{Q}_L \, \Sigma_3 \, \hat{t}_R + 
     \lambda_b \overline{Q}_L \, \hat{\Sigma}_{23} \, b_R +  {\rm h.c.},
 \end{equation}
 where $ \hat{\Sigma}_{23}^i=\epsilon^{ijk} (\Sigma_2^*)_j (\Sigma_3^*)_k$ and $i,j,k=1,2,3$. 
  
Expanding Eq.~(\ref{eq:yukawa}) in component fields we identify the light and the 
heavy right-handed mass eigenstates:
\begin{equation}
  t_R=\frac{1}{\sqrt{\lambda_1^2 f_1^2 +\lambda_2^2 f_2^2}}(\lambda_1 f_1 \hat{t}_R -
            \lambda_2 f_2 \hat{\chi}_R)
    \; \; {\rm and} \; \;  
    \chi_R= \frac{1}{\sqrt{\lambda_1^2 f_1^2+\lambda_2^2 f_2^2}} ( \lambda_2 f_2 \hat{t}_R +
            \lambda_1 f_1\hat{\chi}_R).
\end{equation}
In terms of these mass eigenstates, the Yukawa interactions are
\begin{equation}
 {\cal L}_{yuk}=\sqrt{\lambda_1^2 f_1^2+\lambda_2^2 f_2^2} \, \overline{\chi}_L \chi_R - 
   i \frac{\lambda_t }{\sqrt{2}}(\overline{t}_L,\overline{b}_L) h_2 t_R - 
   i \frac{\lambda_b}{\sqrt{2}} (\overline{t}_L,\overline{b}_L) \hat{h}_2 b_R  + \ldots+  {\rm h.c.},
\end{equation}
where $\hat{h}_2 =  i \sigma_2 h_2^*$ and 
\begin{equation}
\lambda_t =\frac{\lambda_1 \lambda_2 f }{\sqrt{\lambda_1^2 f_1^2 + \lambda_2^2 f_2^2}}.
\end{equation}
Note that the Yukawa coupling of the $b$ quark could have involved $\hat{h}_1$ instead of $\hat{h}_2$. This can be accomplished by replacing $ \hat{\Sigma}_{23}$ by
$ \hat{\Sigma}_{12}$ in the last term in Eq.~(\ref{eq:yukawa}). Therefore, the model can
be either type 1 or type 2 two-Higgs doublet model~\cite{hunter}.

Cancellation of the quadratic divergences introduced by the top quark against the 
divergences  introduced by  the heavy fermion $\chi$ can be verified explicitly.
However, the global symmetries of the Yukawa couplings make the absence of quadratic
divergences apparent.
The terms proportional to $\lambda_1$ and $\lambda_2$ in Eq.~(\ref{eq:yukawa})
separately preserve different $SU(6)\times SU(3)$ subgroups of the global $SU(9)$
symmetry. The $SU(6)\times SU(3)$ subgroups are identical to those subgroups
preserved by the two terms in Eq.~(\ref{eq:stabilize}), and are broken to $SU(5)\times SU(2)$
by $\hat{v}$.

Note that the mass of the heavy partner of the top quark, $\sqrt{\lambda_1^2 f_1^2+\lambda_2^2 f_2^2}$, is not uniquely determined in terms of $\lambda_t$ and $f$. Depending
on the values of  $\frac{f_1}{f_2}$ and $\frac{\lambda_1}{\lambda_2}$ one can alter
the ratio of the $\chi$ mass to the top mass. This will be important in the next section,
where we obtain a lower bound on $f$. Even with $f$ bounded, $\chi$ can be light enough
to promptly cancel quadratic divergences induced by the top quark.
It is quite generic that models with several parameters will have the freedom of varying $m_\chi/m_{top}$ and this feature is also present in the model of Ref.~\cite{KS}.

It is easy to also incorporate lepton Yukawa couplings. We add a vector-like pair of
heavy leptons in order to get a triplet of left-handed fields $L_L=(\nu_L,\tau_L, \psi_L)^T$
with $X$ charge $1/3$. The right-handed fields $\tau_R$ and $\psi_R$ carry charges 1 and 0,
respectively. The Yukawa couplings
\begin{equation}
 {\cal L}_{yuk}= \lambda_2 \bar{L}_L \Sigma_2 \psi_R + 
                \lambda_{\tau} \bar{L}_L \hat{\Sigma}_{23} \tau_R +  {\rm h.c.} 
\end{equation}
give mass of order $f$ to the heavy lepton and a standard Yukawa coupling for the $\tau$.                

\subsection{Electroweak symmetry breaking}
\label{sec:esb}
%%%%%%%%%%%%%%%%%%%%%%%%%%%%%%%%%%%%%%%%%%%%%%%%

The quartic potential displayed in  Eq.~(\ref{eq:quarticfinal}) does not stabilize arbitrary Higgs
VEVs. It leaves a flat directions along which $\langle h_1\rangle \,  \langle h_2\rangle=0$.
This feature is identical to the Higgs potential discussed in Ref.~\cite{SU6Sp6}.
These flat directions are not problematic if  the Higgs potential, in addition to the quartic term,  includes positive Higgs masses and a negative B-term
\begin{equation}
 V_{brk}=m_1^2 h_1^\dagger h_1 + m_2^2 h_2^\dagger h_2+ 
   B (h_1^\dagger h_2 + h_2^\dagger h_1),
\end{equation}
where $m_1^2,m_2^2 >0$, $B<0$. Electroweak symmetry breaking requires
 $B^2 > m_1^2 m_2^2$.

The quartic term leaves additional flat directions  where $Re(h_1)=0$ and $Im(h_2)=0$,
and an analogous flat direction with  $h_1$ and $h_2$ interchanged, because the quartic
term vanishes for $Im(h_1^\dagger h_2)=0$. These additional flat directions are not dangerous
if all coefficients of the Higgs potential are real. This can be accomplished by imposing CP
conservation in the Higgs sector. If imposing CP symmetry is for some reason undesirable
it is easy to modify the terms in Eq.~(\ref{eq:quartic}) to yield
$V_{eff} \propto | h_2^\dagger h_1 |^2.$
One possibility is to modify the matrices $M_{1,2}$ defined in Eq.~(\ref{eq:matricesM})
by erasing one of the nonzero entries in each matrix.
For simplicity, we will assume that all coefficients in the Higgs potential are real.
 
So far, all the interactions we have introduced preserve enough global symmetries to give
two massless Higgs doublets.  The exact symmetries are either an $SU(3)^3$ broken
to $SU(2)^3$ or an $SU(6) \times SU(3)$ broken to $SU(5)\times SU(2)$.
The same set of symmetries is enjoyed by operators that are induced from the tree-level
Lagrangian by one-loop quadratically  divergent diagrams. Therefore, none of these terms can generate a Higgs mass terms. It is often not apparent that the doublet  mass terms are not generated
by the ``Higgs-friendly" operators. When expanding such operators in terms of component
fields, there exist  terms quadratic in Higgs fields accompanied by terms linear in the $SU(2)$ singlet fields. As we discussed at the end of Sec.~\ref{sec:quartic}, one needs to identify the
proper vacuum, in which the tadpoles vanish and the Higgs mass terms are exactly zero.

It is therefore necessary to generate operators with less symmetry. Such operators can
be included in the tree-level Lagrangian with small coefficients. Introducing new
tree-level operators may not be necessary because the logarithmically
divergent one-loop diagrams, as well as two-loop quadratically divergent diagrams,
produce operators with less global symmetry. This is because such diagrams involve
insertions of several operators each with a different symmetry structure. The diagrams involving several operators  generically yield operators that preserve only a common subgroup
of symmetries of all operators involved. The Higgs masses generated by
the log-divergent diagrams and the two-loop quadratically-divergent diagrams are of order $\frac{f}{4 \pi}$, which is the desired order of magnitude.

For example, the log-divergent contribution coming from gauge interactions,
Eq.~(\ref{eq:gaugelog}), gives identical masses terms for $h_1$ and $h_2$.
Similarly, there is a log-divergent contribution from the top quark and its heavy partner $\chi$,
which is proportional to $|\Sigma_3^\dagger \Sigma_2|^2$. This operator also generates
Higgs masses.
 
%%%%%%%%%%%%%%%%%%%%%%%%%%%%%%%%%%%%%%%%%%%%%%%%
\section{Precision Electroweak Measurements}
\label{sec:pheno}
%%%%%%%%%%%%%%%%%%%%%%%%%%%%%%%%%%%%%%%%%%%%%%%%
As is well known \cite{littleprecision1,littleprecision2,SandT,Peskin}
precision electroweak measurements
can place tight constraints on new theories of electroweak symmetry breaking. The simplest
method for calculating these constraints is to construct a series of effective theories by sequentially
integrating out the massive modes.  In the model considered here, 
the $\Sigma$ VEV breaks $SU(3)\times U(1)$ down to $SU(2)_L \times U(1)_Y$ and  gives masses of order $f$ to some of the gauge 
bosons.  With no Higgs VEVs
the mass matrix is diagonalized in the basis:
\beq
B^\mu &=& \frac{1}{\sqrt{g^2 + g_1^2/3}}(g B^\mu_L + \frac{g_1}{\sqrt{3}} B^\mu_H )\, ,\\
  A^\mu_8 &=& \frac{1}{\sqrt{g^2 + g_1^2/3}}(g B^\mu_H - \frac{g_1}{\sqrt{3}}  B^\mu_L)\,  .
  \eeq
The gauge bosons $A^\mu_i$ (for $i=4,5,6,7$) and $B^\mu_H$ get masses 
\beq
M_A &=& \frac{g \,f}{\sqrt{2}}\, ,\\
M_{B_H} &=& \frac{\sqrt{2(3 g^2+g_1^2)} \,f}{3}\, .
\eeq

The Higgs VEVs introduce the usual non-linear sigma model masses for the light
$SU(2)_L \times U(1)_Y$ gauge bosons as well as (in the current model) mixing only between $B^\mu_H$ and $B^\mu_L$. 
Integrating out the heavy gauge bosons and plugging in the Higgs VEVs
\beq
\langle h_1 \rangle & =& \cos \beta\left( \begin{array}{c} v \\ 0 \end{array}\right)\, ,\\
\langle h_2 \rangle &=& \sin \beta\left( \begin{array}{c} v \\ 0 \end{array}\right)
\eeq
we obtain the $W$ and $Z$ masses
\beq
M_W^2&=&\frac{g^2 \,v^2}{4}\left( 1 - \frac{v^2}{6\,f^2} \right)\, ,\\
M_Z^2&=&\frac{(g^2+g^{\prime 2})\, v^2}{4} \left( 1 - \frac{\left( 7\,g^4 - 6\,g^2\,g^{\prime 2} + 3\,g^{\prime 4} \right) \,
     v^2}{24\,f^2\,g^4} \right)\, .
     \eeq
The low-energy neutral current Lagrangian is given by:
  \beq
  {\cal L}_{nc}&= &e A_\mu J^\mu_Q+ 
    \frac{e}{c_W s_W} Z_\mu \left[ 
   J^\mu_3 \left(1+ \frac{ g^{\prime 2}( g^2 -g^{\prime 2}) v^2}{8 g^4 f^2} \right)
    +J^\mu_8   \frac{\sqrt{3} (g^{\prime 2}-g^2)v^2}{8 g^2 f^2} \right. \nonumber \\
 && \left.   -s_W^2 J^\mu_Q
     \left(1+ \frac{(g^{\prime 4}- g^4 -g^{\prime 4}) v^2}{8 g^4 f^2 } \right)
   \right]  -\frac{ (3 g^{\prime 2}-g^2)^2}{2 g^4 f^2} (J_3-J_Q)^2 \nonumber\\
   &&-\frac{g^{\prime 2}(3g^6-2 g^4 g^{\prime 2})v^2}{2 \sqrt{3} g^8 f^2}J_8(J_3-J_Q)-\frac{9 g^4 -6 g^2 g^{\prime 2}+2 g^{\prime 4}}{12 g^4 f^2} J_8^2\, .
  \label{Lnc1}
  \eeq
Note that 
using the relation $J^\mu_8=-\sqrt{3}(J^\mu_Y+J^\mu_B)$ this Lagrangian can be rewritten as:
  \beq
  {\cal L}_{nc}&= &e A_\mu J^\mu_Q+ 
    \frac{e}{c_W s_W} Z_\mu \left[ 
   J^\mu_3 \left(1+ \frac{(4 g^2 g^{\prime 2}-3 g^4 -g^{\prime 4}) v^2}{8 g^4 f^2} \right)
    +J^\mu_X   \frac{3 (g^2-g^{\prime 2})v^2}{8 g^2 f^2} \right. \nonumber \\
 && \left.   -s_W^2 J^\mu_Q
     \left(1+ \frac{(4 g^2 g^{\prime 2}-3 g^4 -g^{\prime 4}) v^2}{8 g^4 f^2 s_W^2} \right)
   \right]  -\frac{ (3 g^{\prime 2}-g^2)^2}{4 g^4 f^2} (J_3-J_Q)^2 -\frac{9}{4 f^2} J_X^2\, .
  \label{Lnc}
  \eeq

It is understood that $J_X$ in the neutral current Lagrangian is replaced in terms of
the quark and lepton currents and does not include the heavy fermion contributions.
The low-energy observables in neutral current scattering are then obtained by
integrating out the $Z$ \cite{littleprecision2}. In addition, the heavy fermions $\chi_{L,R}$ introduced in Sec.~\ref{sec:yukawa} also affect the electroweak couplings of up-type
quarks and  neutrinos.  The
mass mixing of the charge 2/3 quarks can be diagonalized by a bi-unitary transformation,
 and since this is a $2 \times 2$ matrix we can parameterize it by one mixing angle given by
\beq
\theta_{\rm mix}=\frac{v}{f}\eta~.
\eeq
Using Eq. (\ref{eq:yukawa}) in the massless quark limit ($\lambda_1 \rightarrow 0$) the parameter $\eta$ is given by
\beq
\eta=\frac{\sin \beta}{\sqrt{2} }\, \frac{f_1}{f_2}~.
\eeq
Similary by taking  $\lambda_2 \rightarrow 0$ we would arrive at a result for $\eta$ with $f_1$ and $f_2$ interchanged. The result for the neutrinos is identical. Thus the coupling of the up-type quarks and neutrinos to $W^3$ receives a correction factor $ 1- \frac{v^2}{f^2}\eta^2$, while the effective hypercharge of the up-type quarks becomes $\frac{1}{6}+ \frac{v^2}{2f^2}\eta^2$ and the effective hypercharge of the neutrinos becomes $ \frac{-1}{2}+ \frac{v^2}{2f^2}\eta^2$.
There are similar corrections to the couplings to 
 $W^\pm$. The $J_Q$ and $J_X$ currents receive no corrections since
 the mixing is between two fermions
with identical charges under the corresponding gauge symmetries.
   The neutrino mixing affects the value of $G_F$ (in addition  to the correction to the $W$ mass):
     \beq 
  G_F&=&\frac{1}{\sqrt{2} v^2} \left(1 + \frac{v^2}{6\,f^2}- \frac{2 \eta v^2}{f^2}\right) \, .
  \eeq
  
   To compare with experiment we use $\alpha(M_z)$, $G_F$, and $M_Z$ as input parameters,
and the standard definition of the weak mixing angle 
$\sin \theta_0$ from the $Z$ pole \cite{Peskin} 
\beq
\label{s2}
\sin^2 \theta_0 \cos^2 \theta_0 &=& \frac{\pi \alpha(M_Z^2)}{\sqrt{2} G_F M_Z^2}\, ,\\
\sin^2 \theta_0&=&0.23105 \pm 0.00008\, .
\eeq
Expressing the precision electroweak observables in terms of these input parameters
and the corrections to gauge couplings \cite{Burgess}  implied by Eq.~(\ref{Lnc}) 
allows us to express all the deviations of the
observables from SM predictions \cite{Erler} in terms of our model parameters. As can be seen
above the dependence on
$\tan \beta$ drops out of all observables, except for an implicit dependence through $\eta$.
Performing a two parameter fit, we find that weakest bound on $f$ occurs for $\eta \sim 0$, and
is given by $f \ge 3.3$ TeV at the 95\% confidence level as shown in Figure 1. It is clear that the fit to data prefers the region of small mixing. For $\eta=0.1$ the bound grows to $f \ge 3.4$ TeV. It is interesting to note that the bound on $f$ arises mainly from the weak charge of Cesium.  If
this measurement is dropped from the fit the bound goes down to $f \ge 2.5$ TeV.

%%%%%%%%%%%%%%%%%%%%%%%%%%%%%%%%
\begin{figure}[ht]
\centerline{\includegraphics[width=0.4\hsize]{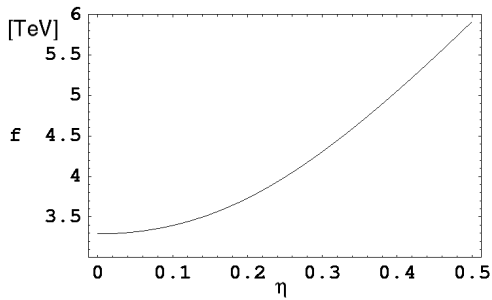}}
\caption[]{95\% confidence level bound on $f$ as a function of the mixing parameter $\eta$.}
\label{fig:f}
\end{figure}
%%%%%%%%%%%%%%%%%%%%%%%%%%%%%%%

The bound $f \ge 3.3$ TeV
corresponds to 
\beq
M_A &\ge& 1.5 \, {\rm TeV}\, ,\\
M_{B_H} &\ge&1.8\,  {\rm TeV} \, .
\eeq
There is no bound on the mass of the heavy fermion $\chi$, since it varies from 0 to $\infty$ over
the parameter space. The region of parameter space where $\chi$ is light is shown in Figure 2
where we have used the constraint $\lambda_t=1$ to eliminate $\lambda_2$.
Plotting the contours in this manner emphasizes large $f_2/f_1$ and large $\lambda_1$, plotting
$f_1/f_2$ versus $\lambda_2$ would emphasize the opposite regime.

%%%%%%%%%%%%%%%%%%%%%%%%%%%%%%%%
\begin{figure}[ht]
\centerline{\includegraphics[width=\hsize]{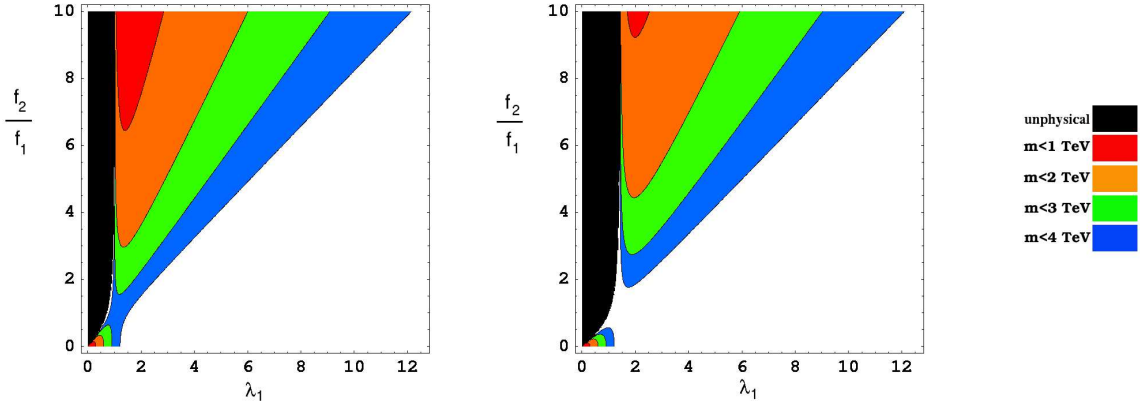}}
\caption[]{Mass of the heavy fermion $\chi$ in the parameter space $f_2/f_1$ versus $\lambda_1$, for large $\tan \beta$ on the left and
$\tan \beta = 1$ on the right.}
\label{fig:mchi}
\end{figure}
%%%%%%%%%%%%%%%%%%%%%%%%%%%%%%%

Since the corrections to the precision electroweak measurements were a factor of two smaller
than those in Ref. \cite{KS} one might have expected that the bound on $f$ would be a factor of  $\sqrt{2}$ weaker, which, given that the bound \cite{littleprecision2} in that case was 4.2 TeV, would yield a bound of 3 TeV.  The discrepancy arises from the fact that the fit for the model of ref. \cite{KS} used four parameters, while in this case we have fit with two parameters and $\delta \chi^2$ grows with the number of fit parameters.

There may also be interesting loop corrections from the Higgs and heavy fermion sectors,
which were recently calculated in \cite{loop} for the model of Ref.~\cite{SU6Sp6}.

%%%%%%%%%%%%%%%%%%%%%%%%%%%%%%%%%%%%%%%%%%%%%%%%
\section{Conclusions}
\label{sec:concl}
%%%%%%%%%%%%%%%%%%%%%%%%%%%%%%%%%%%%%%%%%%%%%%%%

We presented a little Higgs model based on a~simple pattern of global symmetry
breaking and at the same time avoided replicated gauge groups. In many aspects our model
has similar features to the model of Ref.~\cite{KS}, in which the electroweak group
is extended to $SU(4)\times U(1)$. The global symmetry group in our model 
is simple, which might yield more elegant UV completions.

A natural question is about the possibility of  the dynamical origin for the $SU(9)\rightarrow SU(8)$ breaking pattern. While the breaking pattern clearly is not a result of QCD-like dynamics, one can easily conceive of supersymmetric examples. For instance, 9 chiral superfields transforming in  the vector representation of an  $SO(N)$ gauge group will have an $SU(9)$ global symmetry.

We investigated  how the heavy particles affect the Standard Model observables at the
electroweak scale. At 95\% confidence level, the  lower bound on the pion decay constant $f$
is $3.3$~TeV. This corresponds to heavy gauge boson masses of $1.5$ and $1.8$~TeV, which
does not require  fine tuning.  The most significant fine tuning  of the Higgs mass  results from fermion loops, and in little Higgs models the contribution to the Higgs mass from fermion loops
grows with the mass of the heavy partner of the top quark. As we stressed  in Sec.~\ref{sec:yukawa},
the mass of the heavy fermion can be treated as a free parameter
since it depends on the ratios $\frac{\lambda_1}{\lambda_2}$ and $\frac{f_1}{f_2}$.
The ratio $\frac{f_1}{f_2}$ is constrained by the precision electroweak measurements
because it contributes to the quark and neutrino mixings with $SU(2)$ singlet states.
In Sec~\ref{sec:pheno}, we presented  the bounds on the heavy fermion mass resulting
from constraint on $f$.
The heavy fermion can be lighter than 2~TeV in a significant region of the
parameter space and even as light as 1~TeV in small corners of parameter space.

The Higgs sector of our model is similar to those in Refs.~\cite{SU6Sp6,KS}.
The quartic term in the Higgs potential contains only one term, and thus it is far from
being the most general two-Higgs doublet potential. This implies several interesting
relations among the parameters of the Higgs sector~\cite{SU6Sp6}. As in all little Higgs models,
there is a rich spectrum of TeV mass particles: gauge bosons that acquire masses from the breaking
of $SU(3)$ to $SU(2)$, heavy leptons and quarks, and two complex singlet scalars. 
It would be exciting to see experimental confirmation of such particles, but should this happen,
it will be challenging to verify that a set of heavy states comes from a little
Higgs theory~\cite{littlepheno}.
  
%%%%%%%%%%%%%%%%%%%%%%%%%%%%%%%%%%%%%%%%%%%%%%%%
\acknowledgments 
We are grateful to Andy Cohen, David Kosower, and Martin Schmaltz for helpful discussions. 
W.S. is supported in part by the DOE grant DE-FG02-92ER-40704 while J.T. is supported in part
by the US Department of Energy under contract W-7405-ENG-36.
W.S. thanks the theory group of Boston University and the T-8 group at Los Alamos
for their hospitality during enjoyable visits where part of this work was conducted.
J.T. thanks the theory group at Yale for their hospitality during a stimulating visit where part of
this work was done.

%%%%%%%%%%%%%%%%%%%%%%%%%%%%%%%%%%%%%%%%%%%%%%%%
\appendix
\section{Quadratic divergences}
In this appendix we would like to outline how to explicitly check that quadratic divergences do
not generate Higgs mass terms. So far, we used symmetry arguments so it will be reassuring
to see an explicit calculation. We first present a  Feynman diagram computation
and then comment on the Coleman-Weinberg~\cite{CW} effective potential.

 The cancellations of quadratic divergences induced by the loops of
fermions and gauge bosons are straightforward. More interesting
is the cancellation of divergences introduced by the scalar  self interactions. 
In the case of scalar self interactions  there is a subtlety we would like to explain here.
Some of this is certainly known to people working on this topic, but not written in the
literature in any detail.

Let us concentrate on the interactions induced by the potential in Eq.~(\ref{eq:quartic})
and for simplicity set $\kappa_1=\kappa_2=\frac{\kappa}{4}$
so that $ V_q=  \frac{\kappa}{4} \left[(\Sigma^\dagger M_1 \Sigma)^2 +  
(\Sigma^\dagger M_2 \Sigma)^2\right]$. It is most transparent to focus on
one component of the Higgs fields, for example $Re(h_1^d)$. The superscript refers to
the lower component of the Higgs doublet. We expand $V_q$ in components and only
display these terms that could contribute to the $Re(h_1^d)$ mass term 
\begin{equation}
   V_q=\frac{\kappa f^2}{2} (s_1^I)^2 +\frac{\kappa}{12} [Re(h_1^d)]^2 \left(3 [Re(h_2^d)]^2 - 4  (s_1^I)^2 \right)+\ldots
\end{equation}
The diagrams with $Re(h_2^d)$ and $s_1^I$ running in the loops do not cancel against
one another and if this was the whole story they would  generate a quadratically-divergent mass term.
The culprit is the nonlinear kinetic energy term, (\ref{eq:kinetic}), which contains
higher order terms in addition to the ordinary kinetic terms.
\begin{equation}
  {\cal L}_{kin} = \frac{1}{2} (\partial_\mu \pi_i)^2 -\frac{1}{12 f^2} (\partial_\mu s_1^I)^2 [Re(h_1^d)]^2 +\ldots
\end{equation}
The diagram with $s_1^I$ running in the loop with a mass insertion for this field is
also quadratically divergent. It turns out that this diagram cancels the other two
quadratically-divergent diagrams. It might, at first, seem like a coincidence that a diagram
with derivative interactions cancels potential term interactions. However, all three  diagrams
are proportional to the same symmetry-breaking parameter $\kappa$. Moreover, the
normalization of the derivative interaction is uniquely determined by requiring  all fields
have canonical kinetic energy terms.

Similarly, the higher order terms in the kinetic energy Lagrangian prevent an ingenious
approach to computing the Coleman-Weinberg potential. One might be tempted to expand the
$\Sigma$ field around the background fields such that $\Sigma=\exp[ i(\langle \hat{\pi}\rangle+ \hat{\pi})/f]\hat{v}$. The background fields $\langle \hat{\pi}\rangle$ inserted into the higher order
terms in the kinetic energy spoil canonical normalization. A convenient way around this is
to parameterize the nonlinear field as
\begin{equation}
\label{eq:sigmaCW}
 \Sigma =  \exp[ i \langle \hat{\pi} \rangle /f ] \exp[ i  \hat{\pi} /f ]\, \hat{v}
 \end{equation}
 so the background fields drop out of the kinetic energy term.

Using the form (\ref{eq:sigmaCW}) we calculate ${\rm tr}\, {\cal M}^2(\langle \hat{\pi}\rangle)$
keeping $\langle \hat{\pi}\rangle$ to all orders. We get ${\rm tr}\, {\cal M}^2 \propto \kappa [\sin(\langle Re(h_1^d) \rangle)^2 +  \cos(\langle Re(h_1^d) \rangle]^2$,
which is independent of $Re(h_1^d)$. This would not be the case if we used a
parameterization in which the kinetic energy depends on the background fields.

\end{document}